\documentclass{article}
\usepackage{amsfonts,amssymb,amsmath, epsfig}
\usepackage{color,graphicx,graphics,psfrag}
\usepackage{amsmath,amstext,amssymb,amsfonts, amscd}
\usepackage{authblk}

\def\Box{\leavevmode\vbox{\hrule
     \hbox{\vrule\kern4pt\vbox{\kern4pt}%
           \vrule}\hrule}}

\newcounter{appendix}
\def\appendix{\advance\c@appendix by 1
   \def\thesection{\Alph{section}}
   \ifnum\c@appendix=1 \setcounter{section}{-1} \fi
   \@startsection {section}{1}{\z@}{-3.5ex plus -1ex minus 
   -.2ex}{2.3ex plus .2ex}{\Large\bf}}

\def\Section#1{\setcounter{equation}{0} \section{#1} \markboth{#1}{#1}
   \leavevmode\par}
\def\paragraph#1{{\bf #1\ }}

\newtheorem{lemma}{Lemma}[section]  

\newtheorem{theorem}[lemma]{Theorem}

\newtheorem{definition}[lemma]{Definition}

\newtheorem{proposition}[lemma]{Proposition}

\title{Mathematical models of collective dynamics and self-organization} 

\author[(1)]{Pierre Degond}

\affil[(1)]{Department of Mathematics, Imperial College London, 

London, SW7 2AZ, UK,

pdegond@imperial.ac.uk}

\date{} 
\begin{document}

\maketitle

\begin{abstract}
In this paper, we beginning by reviewing a certain number of mathematical challenges posed by the modelling of collective dynamics and self-organization. Then, we focus on two specific problems, first, the derivation of fluid equations from particle dynamics of collective motion and second, the study of phase transitions and the stability of the associated equilibria.  
\end{abstract}

\medskip
\noindent
{\bf Acknowledgements:} PD acknowledges support by the Engineering and Physical Sciences 
Research Council (EPSRC) under grants no. EP/M006883/1 and EP/P013651/1, by the Royal 
Society and the Wolfson Foundation through a Royal Society Wolfson  Research Merit Award no. WM130048 and by the National Science  Foundation (NSF) under grant no. RNMS11-07444 (KI-Net). PD is on leave from CNRS, Institut de Math\'ematiques de Toulouse, France. Works mentionned in this article has been realized in collaboration with many people. I wish to acknowledge more particularly A. Frouvelle, J-G. Liu, S. Merino-Aceituno, S. Motsch and A. Trescases for their decisive contributions.

\medskip
\noindent
{\bf Data statement: } No new data were collected in the course of this research. 

\medskip
\noindent
{\bf Conflict of interest:} The authors declare that they have no conflict of interest. 



\vskip 0.4cm

\Section{Overview}
\label{sec:intro}

Fascinating examples of collective motion can be observed in nature, such as insect swarms \cite{Bazazi_etal_CurrBiol08, Khuong_etal_ECAL11}, bird flocks \cite{Lukeman_etal_PNAS10}, fish schools \cite{Aoki_BullJapSocSciFish92, Degond_Motsch_JSP08, Degond_Motsch_JSP11, Domeier_Colin_BullMarSci97, Gautrais_etal_JMB09, Gautrais_etal_PlosCB12}, or in social phenomena, such as the spontaneous formation of lanes in pedestrian crowds \cite{Moussaid_etal_PlosCB12}. Similarly, at the microscopic scale, collective bacterial migration is frequently observed \cite{Czirok_etal_PRE96} and collective cell migration occurs during organism development \cite{Shraiman_PNAS05} or healing \cite{Poujade_etal_PNAS07}. Such systems of many autonomous agents locally interacting with each other are able to generate large-scale structures of sizes considerably exceeding the perception range of the agents. These large-scale structures are not directly encoded in the interaction rules between the individuals, which are usually fairly simple. They spontaneously emerge when a large number of individuals collectively interact \cite{Vicsek_Zafeiris_PhysRep12}. This is referred to as ``emergence''. 

Emergence is a sort of bifurcation, or phase transition. In physics, phase transitions are dramatic changes of the system state consecutive to very small changes of some parameters, such as the temperature. In self-organized systems, the role of temperature is played by the noise level associated to the random component of the motion of the agents. For instance, in road traffic, the presence of drivers with erratic behavior can induce the formation of stop-and-go waves leading to a transition from fluid to congested traffic. Here, an increase of temperature (the random behavior of some agents) leads to a sudden blockage of the system. This is an example of the so-called ``freezing-by-heating'' phenomenon \cite{Helbing_Farkas_Vicsek_PRL00} also observed in pedestrian crowds and a signature of the paradoxical and unconventional behavior of self-organized systems.

Another parameter which may induce phase transitions is the density of individuals. An increase of this density is very often associated with an increase of the order of the system \cite{Vicsek_etal_PRL95}. For instance, the spontaneous lane formation in pedestrian crowds only appears when the density is high enough. This increase of order with the density is another paradoxical phenomenon in marked contrast with what is observed in more classical physical systems where an increase of density is generally associated with an increase of temperature, i.e. of disorder (this can be observed when pumping air into a bicycle tire: after using it, the pump core has heated up). 

The passage between two different phases is called a critical state. In physical systems, critical states appear only for well-chosen ranges of parameters. For instance, at ambient pressure, liquid water passes to the gaseous state at the temperature of $100$ $^\circ$C. In self-organized systems, by contrast, critical states are extremely robust: they appear almost systematically, whatever the initial conditions of the system. In dynamical systems terms, the critical state is an attractor. The presence of critical states which are attractors of the dynamics is called ``Self-Organized Criticality'' \cite{Bak_etal_PRL87} and its study is important in physics. 

We shall focus on models of collective dynamics and self-organization that provide a prediction from an initial state of the system. These are stated as Cauchy problems for appropriate systems of differential equations. The modelling of self-organization meets important scientific and societal challenges. There are environmental and societal stakes: for instance, better understanding the behavior of a gregarious species can lead to improved conservation policies ; modelling human crowds improves the security, efficiency and profitability of public areas ; understanding collective cell migration opens new paradigms in cancer treatment or regenerative medicine. There are also technological stakes: roboticists use social interaction mechanisms to geer fleets of robots or drones ; architects study social insect nests to look for new sustainable architecture ideas. 

Large systems of interacting agents (aka particles) are modelled at different levels of detail. The most detailed models are particle models (aka individual-based or agent-based models). They describe the position and state of any single agent (particle) of the system as it evolves in time through its interactions with the other agents and the environment. This leads to large coupled systems of ordinary or stochastic differential equations (see an example in \cite{Vicsek_etal_PRL95}). When the number of particles is large, these systems are computationally intensive as their cost increases polynomially with the number of particles. Additionally their output is not directly exploitable as we are only interested in statistical averages (e.g. the pressure in a gas) and requires some post-processing which can generate errors. 

For this reason, continuum models are often preferred \cite{Toner_etal_PRE98}. They consist of partial differential equations for averaged quantities such as the mean density or mean velocity of the agents. However, in the literature, a rigorous and systematic link between particle and continuum models is rarely found. Yet, establishing such a link is important. Indeed, often, the microscopic behavior of the agents is not well-known and is the actual target. On the other hand, large-scale structures are more easily accessible to experiments and can be used to calibrate continuum models. But to uncover the underlying individual behavior requires the establishment of a rigorous correspondence between the two types of models. Our goal is precisely to provide methodologies to establish this correspondence. 

To derive continuum models from particle models rigorously requires a coarse-graining methodology. There are two steps of coarsening. The first step consists of deriving a ``kinetic model'', which provides the time evolution of the probability distribution of the agents in position and state spaces. The equation for this kinetic distribution can be derived from the particle model, however not in closed form unless one assumes a strong hypothesis named ``propagation of chaos'' which means statistical independence between the particles. This hypothesis is generally wrong but admittedly,  becomes asymptotically valid as the particle number tends to infinity. To prove such a result is a very difficult task and until recently \cite{Gallagher_etal_EMS13, Mischler_Mouhot_InvMath13}, the only available result one was due to Lanford for the Boltzmann model \cite{Lanford_76}. Kinetic models are differential or integro-differential equations posed on a large dimensional space such as the Boltzmann or Fokker-Planck equations.

The second step of coarsening consists of reducing the description of the system to a few macroscopic averages (or moments) such as the density or the mean velocity as functions of position and time. The resulting fluid models are systems of nonlinear partial differential equations such as the Euler or Navier-Stokes equations. Fluid models are derived by averaging out the state variable of kinetic models (such as the particle velocity) to only keep track of the spatio-temporal dependence. Here again, a closure assumption is needed, by which one postulates a known shape of the distribution function as functions of its fluid moments. It can be justified in the hydrodynamic regime when the kinetic phenomena precisely bring the distribution function close to the postulated one. Providing a rigorous framework to these approaches is the core subject of ``kinetic theory'', whose birthdate is the statement of his 6th problem by Hilbert in his 1900 ICM address. Since then, kinetic theory has undergone impressive developments, with Field's medals awarded to P. L. Lions and C. Villani for works in this theory.

It is therefore appealing to apply kinetic theory methods to collective dynamics and self-organization. However, this has proved more delicate than anticipated and fascinating new mathematical questions have emerged from these difficulties. A first difficulty is that kinetic models may lose validity as propagation of chaos may simply be not true. Indeed, self-organization supposes the build-up of correlations between the particles. It is not clear that these correlations disappear with the number of particles tending to infinity. We have indeed proved (with E. Carlen and B. Wennberg \cite{Carlen_etal_M3AS13}) in a simple collective dynamics model that propagation of chaos may break down at large temporal scales. Are there new models that can replace the defective kinetic equations when propagation of chaos breaks down ? Some phenomenological answers have been proposed but to the best of our knowledge, no mathematical theory is available yet.

A second difficulty arises at the passage between kinetic and fluid models. In classical physics, a fundamental concept is that of conservation law (such as mass, momentum or energy conservations). These conservation laws are satisfied at particle level and so, are transferred to the macroscopic scale and serve as corner stone in the derivation of fluid equations. By contrast, biological or social systems are open systems which exchange momentum and energy with the outside world and have no reason to satisfy such conservation laws. This is a major difficulties as acknowledged in Vicsek's review \cite{Vicsek_Zafeiris_PhysRep12}. In a series of works initiated in \cite{Degond_Motsch_M3AS08}, we have overcome this problem and shown that some weaker conservation laws which we named ``generalized collision invariants (GCI)'' prevail. They enabled us to derive fluid  models showing new and intringuing properties. Their mathematical study is still mostly open.  We will provide more details in Section~\ref{sec:fluid}.

The third difficulty is linked to the ubiquity of phase transitions in self-organized systems. This puts a strong constraint on fluid models which must be able to correctly describe the various phases and their interfaces. Complex phenomena like hysteresis \cite{Couzin_etal_JTB02}, which results from the presence of multiple stable equilibria and involves the time-history of the system, must also be correctly rendered. However, different phases are described by types of fluid models. For instance, in symmetry-breaking phase transitions, the disordered phase is described by a parabolic equation while the ordered phase is described by a hyperbolic equation \cite{Degond_etal_JNonlinearSci13, Degond_etal_arXiv:1304.2929}. At the critical state, these two phases co-exist and should be related by transmission conditions through phase boundaries. These transmission conditions are still unknown. More about phase transitions can be found in Section~\ref{sec:phasetrans} and references \cite{Barbaro_Degond_DCDSB13, Frouvelle_Liu_SIMA12}. Convergence to swarming states for the Cucker-Smale model \cite{Cucker_Smale_IEEETransAutCont07} has been extensively studied in the mathematical literature \cite{Carrillo_etal_SIMA10, Ha_Liu_CMS09, Ha_Tadmor_KRM08, Motsch_Tadmor_JSP11, Shen_SIAP07}, as well as for related models \cite{Chuang_etal_PhysicaD07}. 

We have used symmetry-breaking phase transitions in a surprising context: to design automatized fertility tests for ovine sperm samples \cite{Creppy_etal_Interface16}. Other types of phase transition play important roles. One of them is the packing transition which occurs when finite size particles reach densities at which they are in contact with each other. This transition occurs for instance in cancer tumors \cite{Leroy_etal_BMB17}, crowds \cite{Degond_Hua_JCP13, Degond_etal_JCP11}, road traffic \cite{Berthelin_etal_ARMA08}, herds \cite{Degond_etal_JSP10} or tissue self-organization \cite{Peurichard_etal_JTB17}. Another example is the transition from a continuum to a network, and is at play for instance in the emergence of ant-trail networks \cite{Boissard_etal_JMB13, Haskovec_etal_NonlinAnalTMA16}. For such systems, many challenges remain such as the derivation of macroscopic models. 

In the forthcoming sections, we will focus on two specific aspects: the derivation of fluid models in spite of the lack of conservations relations (Section \ref{sec:fluid}) and the investigation of phase transitions (Section \ref{sec:phasetrans}).


\section{Derivation of fluid models}
\label{sec:fluid}

\subsection{The Vicsek model}
\label{subsec:vicsek}

We start with the description of particle models of collective behavior. As an example, we introduce the Vicsek model \cite{Vicsek_etal_PRL95} (see related models in \cite{Bertin_etal_JPhysA09, Degond_etal_DCDSB16, Ginelli_etal_PRL10}). It considers systems of self-propelled particles moving with constant speed (here supposed equal to $1$ for notational simplicity) and interacting with their neighbors through local alignment. Such a model describes the dynamics of bird flocks and fish schools  \cite{Vicsek_Zafeiris_PhysRep12}. It is written in the form of the following stochastic differential system: 
\begin{eqnarray}
&& \hspace{-1cm}
dX_i(t) = V_i(t) dt, \label{eq:vicsek1} \\
&& \hspace{-1cm}
dV_i(t) = P_{V_i(t)^\bot} \circ (F_i(t) \, dt + \sqrt{2 \, \tau} dB_t^i), \label{eq:vicsek2} \\
&& \hspace{-1cm}
F_i(t) = \nu \, U_i(t), \quad U_i(t) = \frac{J_i(t)}{|J_i(t)|}, \quad J_i(t) = \sum_{j \, | \,  |X_j(t) - X_i(t)|\leq R} V_j(t). \label{eq:vicsek3}
\end{eqnarray}
Here, $X_i(t) \in {\mathbb R}^d$ is the position of the $i$-th particle (with $i \in \{1, \ldots, N\}$), $V_i(t) \in {\mathbb S}^{d-1}$ is its velocity direction. $B_t^i$ are standard independent Brownian motions in ${\mathbb R}^d$ describing idiosyncratic noise i.e. noise specific to each agent and $\sqrt{2 \, \tau}$ is a constant and uniform noise intensity. $F_i$ is the alignment force acting on the particles: it is proportional to the mean orientation $U_i(t) \in {\mathbb S}^{d-1}$ of the agents around agent $i$, with a constant and uniform multiplication factor $\nu$ encoding the alignment force intensity. $U_i(t)$ itself is obtained by normalizing the total momentum $J_i(t)$ of the agents belonging to a ball of radius $R$ centered at the position $X_i(t)$ of agent $i$. The normalization of $J_i(t)$ (i.e. its division by $|J_i(t)|$ where $|\cdot|$ denotes the euclidean norm) makes only sense if $J_i(t) \not = 0$, which we assume here. The projection $P_{V_i(t)^\bot}$ onto $\{V_i(t)\}^\bot$ is there to maintain $V_i(t)$ of unit norm and is a matrix given by $P_{V_i^\bot} = \mbox{Id} - V_i \otimes V_i$ where Id is the identity matrix of ${\mathbb R}^d$ and $\otimes$ denotes the tensor product. The Stochastic Differential Equation (\ref{eq:vicsek2}) is understood in the Stratonovich sense, hence the symbol $\circ$, so that the noise term provides a Brownian motion on the sphere ${\mathbb S}^{d-1}$ \cite{Hsu_AMS02}. Eq. (\ref{eq:vicsek2}) models two antagonist effects acting on the particles: the alignment force (the first term) which has a focusing effect and the noise (the second term) which has a defocusing effect. The original model proposed in \cite{Vicsek_etal_PRL95} is a time-discretized variant of this model. 

Next, we present the kinetic model corresponding to this discrete model. It is written: 
\begin{eqnarray}
&& \hspace{-1cm}
\partial_t f + \nabla_x \cdot (vf) = \nabla_v \cdot \big( - (P_{v^\bot} \, F_f) \, f + \tau \, \nabla_v f \big), \label{eq:vicsek_KM1} \\
&& \hspace{-1cm}
F_f(x,t) = \nu \, U_f(x,t), \quad U_f(x,t) = \frac{J_f(x,t)}{|J_f(x,t)|}, \label{eq:vicsek_KM2} \\
&& \hspace{-1cm}
J_f(x,t) = \int_{|y-x| \leq R} \int_{{\mathbb S}^{d-1}} f(y,w,t) \, w \, dw \, dy, \label{eq:vicsek_KM3} 
\end{eqnarray}
where $f=f(x,v,t)$ is the particle distribution function and is a function of the position $x \in {\mathbb R}^d$, velocity $v \in {\mathbb S}^{d-1}$ and time $t>0$, $\nabla_v$ stands for the nabla operator on the sphere ${\mathbb S}^{d-1}$ and $P_{v^\bot}$ is the projection operator on $\{v\}^\bot$. $f(x,v,t)$ represents the probability density of particles in the $(x,v)$ space. The left-hand side of (\ref{eq:vicsek_KM1}) describes motion  of the particles in physical space with speed $v$, while the right-hand side models the contributions of the alignment force $F_f$ and of velocity diffusion (with diffusion coefficient $\tau$) induced by Brownian noise at the particle level. The construction of the force term follows the same principles as for the discrete model, with $F_f(x,t)$, $U_f(x,t)$, $J_f(x,t)$ replacing $F_i(t)$, $U_i(t)$, $J_i(t)$. The sum of the velocities over neighboring particles in the computation of the momentum (\ref{eq:vicsek3}) is replaced by  integrals of the velocity weighted by $f$, with spatial integration domain being the ball centered at $x$ and of radius $R$, and velocity integration domain being the whole sphere ${\mathbb S}^{d-1}$ (Eq. (\ref{eq:vicsek_KM3})). Analysis of this model can be found in \cite{Figalli_etal_arxiv15, Gamba_Kang_ARMA16}. The passage from (\ref{eq:vicsek1})-(\ref{eq:vicsek3}) to (\ref{eq:vicsek_KM1})-(\ref{eq:vicsek_KM2}) is shown in~\cite{Bolley_etal_AML11}, in the variant  where $J_i(t)$ is directly used in (\ref{eq:vicsek2}) instead of $F_i(t)$. In the case presented here, the control of $J_i(t)$ away from zero presents additional difficulties which haven't been solved yet. 

The macroscopic equations describe a large spatio-temporal scale regime. This regime is modelled by a time and space rescaling in (\ref{eq:vicsek_KM1})-(\ref{eq:vicsek_KM2}) involving a small parameter $\varepsilon \ll 1$ describing the ratio between the micro and the macro scales,  which leads to
\begin{eqnarray}
&& \varepsilon \big( \partial_t f^\varepsilon + \nabla_x \cdot (vf^\varepsilon) \big) = \nabla_v \cdot \big( - (P_{v^\bot} \, F_{f^\varepsilon}) \, f^\varepsilon + \tau \, \nabla_v f^\varepsilon \big), \label{eq:vicsek_KM1_eps} \\
&& F_f (x,t) =\nu \, u_f(x,t), \quad u_f(x,t)=\frac{j_f(x,t)}{|j_f(x,t)|},\label{eq:vicsek_KM2_eps} \\ 
&& j_f(x,t) = \int_{{\mathbb S}^{d-1}} f(x,w,t) \, w \, dw. \label{eq:vicsek_KM3_eps} 
\end{eqnarray}
The scale change brings a factor $\varepsilon$ in front of the terms at the left-hand side of (\ref{eq:vicsek_KM1_eps}) describing the motion of the particles in position space. It also localizes the integral describing the momentum of particles which now only involves an integration with respect to the velocity $w$ of the distribution at the same location $x$ as the particle onto which the force applies (see Eq. (\ref{eq:vicsek_KM2_eps})). This is due to the interaction radius $R$ being of order $\varepsilon$ in this regime. The expansion of $J_f$ in powers of $\varepsilon$ leads to (\ref{eq:vicsek_KM2_eps}) up to terms of order $\varepsilon^2$ which are neglected here as not contributing to the final macroscopic model at the end. The macroscopic model is obtained as the limit $\varepsilon \to 0$ of this perturbation problem.

Before stating the result, we introduce the ``von Mises Fisher (VMF)'' distribution of orientation $u$ and concentration parameter $\kappa$ where $u$ is an arbitrary vector in ${\mathbb S}^{d-1}$ and $\kappa \in [0,\infty)$. This distribution denoted by $M_{\kappa u}$ is such that for all $v \in {\mathbb S}^{d-1}$: 
\begin{equation}
M_{\kappa u} (v) = \frac{1}{Z} \exp \big( \kappa \, u \cdot v \big), 
\label{eq:vmf}
\end{equation}
where $u \cdot v$ is the euclidean inner product of $u$ and $v$ and $Z$ is a normalization constant only depending on $\kappa$. 
In \cite{Degond_Motsch_M3AS08}, we proved the following formal theorem

\begin{theorem} 
If the solution $f^\varepsilon$ of  (\ref{eq:vicsek_KM1_eps}), (\ref{eq:vicsek_KM2_eps}) has a limit $f^0$ when $\varepsilon \to 0$, it is given by 
\begin{equation}
f^0(x,v,t) = \rho(x,t) \, M_{\kappa u(x,t)} (v), 
\label{eq:vicsek_equi}
\end{equation}
where $\kappa = \nu/\tau$ and the pair $(\rho,u)$ satisfies 
the following ``self-organized hydrodynamic'' (SOH) model: 
\begin{eqnarray}
&& \partial_t \rho + c_1 \nabla_x \cdot (\rho u) = 0, \label{eq:soh1} \\
&& \rho \big( \partial_t u + c_2 (u \cdot \nabla_x u) \big) + \tau \, P_{u^\bot} \nabla_x \rho = 0 , \label{eq:soh2} \\
&& |u|=1, \label{eq:soh3}
\end{eqnarray}
with the coefficients $c_1, c_2$ depending on $\nu$ and $\tau$ and $P_{u^\bot}$ being the projection onto~$\{u\}^\bot$. 
\label{thm_SOH}
\end{theorem}

The VMF distribution provides a way to extend the concept of Gaussian distribution to statistical distributions defined on the sphere. The orientation $u$ describes the mean orientation of the particles while $1/\kappa$ measures the dispersion of the particles around this mean. When $\kappa$ is close to zero, the VMF is close to a uniform distribution while when it is large, it is close to a Dirac delta at $u$. The theorem states that at large scales, the distribution function approaches a VMF distribution weighted by the local density $\rho$. However, both $\rho$ and the orientation $u$ of the VMF depend on position and space and they are determined by solving the SOH model. 

The SOH model is akin to the compressible Euler equations of gas dynamics, but with some important differences. First, the mean orientation $u$ is constrained to lie on the sphere as (\ref{eq:soh3}) shows. The presence of the projection $P_{u^\bot}$ in (\ref{eq:soh2}) guarantees that it is the case as soon as the initial orientation $u|_{t=0}$ belongs to the sphere. The presence of $P_{u^\bot}$ makes the system belong to the class of non-conservative hyperbolic problems, which are notoriously difficult (we can show that the model is hyperbolic). Finally, the convection terms in the two equations are multiplied by different coefficients $c_1 \not = c_2$, while they are the same in standard gas dynamics. This is a signature of a non-Galilean invariant dynamics. Indeed, as the particles are supposed to move with speed $1$, there is a preferred frame in which this speed is measured. In any other Galilean frame this property will be lost. The mathematical properties of the SOH model are open, except for a local existence result in \cite{Degond_etal_MAA13}. A rigorous proof of Theorem \ref{thm_SOH} has been given in \cite{Jiang_etal_arxiv15}.

To understand how Theorem \ref{thm_SOH} can be proved, we write (\ref{eq:vicsek_KM1_eps}) as
\begin{eqnarray}
&& \partial_t f^\varepsilon + \nabla_x \cdot (vf^\varepsilon) = \frac{1}{\varepsilon} Q(f^\varepsilon) \label{eq:vicsek_KM31_eps} \\
&& Q(f) = \nabla_v \cdot \big( - (P_{v^\bot} \, F_f) \, f + \tau \, \nabla_v f \big), \label{eq:vicsek_KM4_eps} 
\end{eqnarray}
with $F_f$ given by (\ref{eq:vicsek_KM2_eps}), (\ref{eq:vicsek_KM3_eps}). It is readily seen that $Q(f)$ can be written as 
\begin{eqnarray}
&& Q(f) = {\mathcal Q} (f;u_f),
\label{eq:QQ} 
\end{eqnarray}
where $u_f$ is the mean orientation associated with $f$ and is given by (\ref{eq:vicsek_KM2_eps}) and where for any $u \in {\mathbb S}^{d-1}$, 
\begin{eqnarray}
&& {\mathcal Q} (f;u) (v) = \tau \, \nabla_v \cdot \Big( M_{\kappa u} (v) \nabla_v \big( \frac{f(v)}{M_{\kappa u} (v)} \big) \Big). 
\label{eq:calQ} 
\end{eqnarray}
We note that for a given $u \in {\mathbb S}^{d-1}$, the operator ${\mathcal Q} (\cdot;u)$ is linear. However, this is not the linearization of $Q$ around $\rho M_{\kappa u}$ as extra terms coming from the variation of $u_f$ with respect to $f$ would appear.  

By formally letting $\varepsilon \to 0$ in (\ref{eq:vicsek_KM31_eps}), we get that $f^0$ is a solution of $Q(f^0)=0$. It is an easy matter to show that this implies the existence of two functions $\rho(x,t)$ and $u(x,t)$ with values in $[0,\infty)$ and ${\mathbb S}^{d-1}$ respectively such that (\ref{eq:vicsek_equi}) holds. Indeed, from (\ref{eq:calQ}) and Green's formula, we get
\begin{eqnarray}
&& \hspace{-1cm}
\int {\mathcal Q} (f;u)(v) \,  \frac{f(v)}{M_{\kappa u} (v)} \, dv  = - d \int M_{\kappa u} (v) \Big| \nabla_v \big( \frac{f(v)}{M_{\kappa u} (v)} \big) \Big|^2 \, dv \leq 0. 
\label{eq:entrop_calQ} 
\end{eqnarray}
Therefore, if ${\mathcal Q} (f;u)=0$, this implies that $\frac{f(v)}{M_{\kappa u} (v)}$ does not depend on $v$. The result follows easily.

To find the equations satisfied by $\rho$ and $u$, it is necessary to remove the $1/\varepsilon$ singularity in (\ref{eq:vicsek_KM31_eps}), i.e. to project the equation on the slow manifold. In gas dynamics, this is done by using the conservations of mass, momentum and energy. Here, the model only enjoys conservation of mass, which is expressed by the fact that 
\begin{eqnarray}
\int Q(f) \,   dv  =  0, \quad \forall f. 
\label{eq:mass_calQ} 
\end{eqnarray}
Hence, integrating (\ref{eq:vicsek_KM31_eps}) with respect to $v$ and using (\ref{eq:mass_calQ}), we get that 
\begin{eqnarray}
&& \partial_t \rho_{f^\varepsilon} + \nabla_x \cdot j_{f^\varepsilon} = 0. \label{eq:mass_cons_1} 
\end{eqnarray}
Letting $\varepsilon \to 0$, with (\ref{eq:vicsek_equi}), we get
\begin{eqnarray}
\rho_{f^\varepsilon} \to \rho, \quad j_{f^\varepsilon} \to j_{f^0} = c_1 \rho u, \label{eq:mass_cons_2} 
\end{eqnarray}
where $c_1$ is the so called order-parameter and is given by 
\begin{eqnarray}
c_1 = c_1(\kappa) = \int M_{\kappa u} (v) \, (v \cdot u) \, dv. 
. \label{eq:order_param} 
\end{eqnarray}
This leads to (\ref{eq:soh1}). 

We need another equation to find $u$. In gas dynamics, this is done by using momentum conservation, which in this context would be expressed by $\int Q(f) \, v \,   dv  =  0$. However, this equation is not true and the lack of momentum conservation relates to the particles being self-propelled and therefore, able to extract or release momentum from the underlying medium. However, in \cite{Degond_Motsch_M3AS08}, I showed that weaker forms of conservations (named generalized collision invariants or GCI) hold and provide the missing equation. 

More precisely, we define
\begin{definition}
For a given orientation $u \in {\mathbb S}^{d-1}$, we define a GCI associated with $u$ as a function $\psi(v) $ such that 
\begin{eqnarray}
\int {\mathcal Q} (f;u)(v) \, \psi(v) \,  dv  =  0, \quad \forall f \, \mbox{ such that } \, P_{u^\bot} j_f =0. 
\label{eq:vel_calQ} 
\end{eqnarray}
\label{def:GCI}
\end{definition}
By restricting the set of $f$ to which we request the conservations to apply, we enlarge the set of candidate GCI $\psi$. In \cite{Degond_Motsch_M3AS08} (see also \cite{Frouvelle_M3AS12}), we show that the following theorem: 

\begin{theorem}
The set ${\mathcal C}_u$ of GCI associated to a given orientation $u$ is a linear vector space of dimension $d$ expressed as follows: 
\begin{eqnarray}
{\mathcal C}_u = \{ C + A \cdot P_{u^\bot} v \, h(u \cdot v) \, \, | \, \, C \in {\mathbb R}, \, \, A \in \{u\}^\bot \}. 
\label{eq:GCI_space} 
\end{eqnarray}
Here, defining $\theta$ by $\cos \theta = u \cdot v$,  $h$ is given by 
\begin{eqnarray}
h(\cos \theta) = \frac{g(\theta)}{\sin \theta}, \quad \theta \in (0,\pi), 
\label{eq:def_h} 
\end{eqnarray}
with $g$ being the unique solution of the elliptic problem 
\begin{eqnarray}
- \frac{d}{d \theta} \Big( \sin^{d-2} \theta \, e^{\kappa \, \cos \theta} \, \frac{dg}{d \theta} \Big) + (d-2) \, \sin^{d-4} \theta \, e^{\kappa \, \cos \theta}  \, g = \sin^{d-1} \theta \, e^{\kappa \, \cos \theta} 
\label{eq:def_g} 
\end{eqnarray}
in the space 
\begin{eqnarray}
V = \{ g \, \, | \, \, (d-2) \, \sin^{\frac{d}{2}-2} \theta \, g \in L^2(0,\pi), \quad \sin^{\frac{d}{2}-1} \theta \, g \in H^1_0(0,\pi) \}.
\label{eq:def_V} 
\end{eqnarray}
We recall that $L^2(0,\pi)$ is the Lebesgue space of square-integrable functions on $(0,\pi)$ and $H^1_0(0,\pi)$ is the Sobolev space of functions which are in $L^2(0,\pi)$ and whose first order derivative is in $L^2(0,\pi)$ and which vanish at $0$ and $\pi$. 
\label{thm:GCI}
\end{theorem}

The GCI have the remarkable property that  
\begin{eqnarray}
\int Q(f) \, P_{u_f^\bot} v \, h(u_f \cdot v) \, dv = 0, \quad \forall f.
\label{eq:cancel} 
\end{eqnarray}
Indeed, $P_{u_f^\bot} v \, h(u_f \cdot v)$ is a GCI $\psi$ associated with $u_f$. Thus, using (\ref{eq:QQ}), and the definition (\ref{eq:vel_calQ}) of GCI, we get
$$ \int Q(f) \, \psi(v) \, dv = \int {\mathcal Q}(f,u_f) \, \psi(v) \, dv =0,$$
as $P_{u_f^\bot} j_f = |j_f| P_{u_f^\bot} u_f =0$. Multiplying (\ref{eq:vicsek_KM31_eps}) by $P_{u_{f^\varepsilon}^\bot} v \, h(u_{f^\varepsilon} \cdot v)$, applying (\ref{eq:cancel}) with $f = f^\varepsilon$ to cancel the right-hand side of the resulting equation, letting $\varepsilon \to 0$ and using (\ref{eq:vicsek_equi}), we get:
\begin{eqnarray}
P_{u^\bot} \, \int \big( \partial_t + v \cdot \nabla_x \big) (\rho \, M_{\kappa u}) \, h(u \cdot v) \, v \, dv  = 0. 
\label{eq:eq_u} 
\end{eqnarray}
After some computations, this equation gives rise to (\ref{eq:soh2}), where the constant $c_2$ depends on a suitable moment of the function $h$. 

The GCI concept has provided a rigorous way to coarse-grain a large class of KM sharing similar structures \cite{Degond_etal_M3AS2016, Degond_etal_DCDSB16, Degond_Motsch_JSP11}. As an example, we now consider the model of \cite{Degond_etal_M3AS2016, Degond_etal_MMS17} where self-propelled agents try to coordinate their full body attitude. This model is described in the next section.

\subsection{A new model of full body attitude alignment}
\label{subsec:body}

The microscopic model considers~$N$ agents with positions~$X_i(t) \in {\mathbb R}^3$ and associated rotation matrices ~$A_i(t) \in \mbox{SO}(3)$ representing the rotation needed to map a fixed reference frame $(e_1, e_2, e_3)$ to the local frame $(A_i(t) \, e_1$, $A_i(t) \, e_2$, $A_i(t) \, e_3)$ attached to the body of agent $i$ at time $t$. As the particle are self-propelled, agent $i$ moves in the direction $A_i(t) \, e_1$ with unit speed.  Agents try to coordinate their body attitude with those of their neighbors. Following these principles, the particle model is written:
\begin{eqnarray}
&& \hspace{-1cm}
dX_i(t) = A_i(t) \, e_1 \,  dt, \label{eq:body1} \\
&& \hspace{-1cm}
dA_i(t) = P_{T_{A_i(t)}} \circ (F_i(t) \, dt + 2 \, \sqrt{\tau} dB_t^i), \quad F_i(t) = \nu \, \Lambda_i(t), \label{eq:body2} \\
&& \hspace{-1cm}
\Lambda_i(t) = \mbox{PD}(G_i(t)), \quad G_i(t) = \sum_{j \, | \,  |X_j(t) - X_i(t)|\leq R} A_j(t). \label{eq:body3}
\end{eqnarray}
Here, $B_t^i$ are standard independent Brownian motions in the linear space of $3 \times 3$ matrices (in which $\mbox{SO}(3)$ is isometrically imbedded) describing idiosyncratic noise and 2 $\sqrt{\tau}$ is the noise intensity. $F_i$ is the force that aligns the body attitude of Agent $i$ to the mean body attitude of the neighbors defined by $\Lambda_i(t)$ with a force intensity $\nu$. $\Lambda_i(t)$ is obtained by normalizing the matrix $G_i(t)$ constructed as the sum of the rotation matrices of the neighbors in a ball of radius $R$ centered at the position $X_i(t)$ of Agent $i$. The normalization is obtained by using the polar decomposition of matrices. We suppose that $G_i(t)$ is non-singular. Then there exists a unique rotation matrix $\mbox{PD}(G_i(t))$ and a unique symmetric matrix $S_i(t)$ such that $G_i(t) = \mbox{PD}(G_i(t)) \,  S_i(t)$. The quantity $P_{T_{A_i(t)}}$ denotes the orthogonal projection onto the tangent space $T_{A_i(t)}$to $\mbox{SO}(3)$ at $A_i(t)$ to guarantee that the dynamics maintains $A_i(t)$ on $\mbox{SO}(3)$. The Stochastic Differential Equation (\ref{eq:body2}) is again understood in the Stratonovich sense, using the symbol $\circ$ to highlight this fact. As a consequence, the noise term provides a Brownian motion on $\mbox{SO}(3)$ as shown in \cite{Hsu_AMS02}. Note however that the noise intensity is $2 \, \sqrt{\tau}$ instead of $\sqrt{2 \tau}$ as before. This is because we endow $\mbox{SO}(3)$ with the inner product $A \cdot B = \frac{1}{2} \mbox{Tr} (A^T B)$, where Tr stands for the trace and the exponent $T$ for the matrix transpose, which corresponds to the standard metric on  $3 \times 3$ matrices divided by $2$. With this convention, the noise $2 \, \sqrt{\tau}$ will exactly yields a diffusion coefficient equal to $\tau$ in the mean-field limit. 

The mean-field model now provides the evolution of the distribution function $f=f(x,A,t)$ which depends on the position $x \in {\mathbb R}^d$, rotation matrix  $A \in \mbox{SO}(3)$ and time $t>0$. It is written
\begin{eqnarray}
&& \hspace{-1cm}
\partial_t f + \nabla_x \cdot (A\, e_1 f) = \nabla_A \cdot \big( - (P_{T_A} \, F_f) \, f + \tau \, \nabla_A f \big), \label{eq:body_KM1} \\
&& \hspace{-1cm}
F_f(x,t) = \nu \, \Lambda_f(x,t), \quad \Lambda_f(x,t) = \mbox{PD} (G_f(x,t)), \label{eq:body_KM2} \\
&& \hspace{-1cm}
G_f(x,t) = \int_{|y-x| \leq R} \int_{\mbox{\footnotesize SO}(3)} f(y,B,t) \, B \, dB \, dy, \label{eq:body_KM3} 
\end{eqnarray}
Here, as pointed out before, $\nabla_A$ and $\nabla_A \cdot$ stand for the gradient and divergence operators on $\mbox{SO}(3)$ when endowed with the Riemannian structure induced by the euclidean norm $\|A\| = \frac{1}{2} \mbox{Tr} (A^T A)$. The measure on $\mbox{SO}(3)$ is the Haar measure normalized to be a probability measure. The passage from (\ref{eq:body1})-(\ref{eq:body3}) to (\ref{eq:body_KM1})-(\ref{eq:body_KM3}) is open but in a variant where $G_i$ is used in the expression of $F_i$ instead of $\Lambda_i$, the proof of \cite{Bolley_etal_AML11} is likely to extend rather straightforwardly. In the case presented here, the control of $G_i(t)$ away from the set of singular matrices presents additional challenges. To the best of our knowledge, the mathematical theory of this model is nonexistent. 

A similar rescaling as in the previous section leads to the following perturbation problem (dropping terms of order $\varepsilon^2$):
\begin{eqnarray}
&& \hspace{-1cm}
\varepsilon \big( \partial_t f^\varepsilon + \nabla_x \cdot (A\, e_1 \, f^\varepsilon) \big) = \nabla_A \cdot \big( - (P_{T_A} \, \, F_{f^\varepsilon}) \, f^\varepsilon + \tau \, \nabla_A f^\varepsilon \big), \label{eq:body_KM1_eps} \\
&& \hspace{-1cm}
 F_f (x,t) =\nu \, \lambda_f(x,t), \quad \lambda_f(x,t)=\mbox{PD} (g_f(x,t)), \label{eq:body_KM2_eps} \\ 
&& \hspace{-1cm}
 g_f(x,t) = \int_{\mbox{\footnotesize SO}(3)} f(x,B,t) \, B \, dB, \label{eq:body_KM3_eps} 
\end{eqnarray}
where we have denoted by $g_f$ the local modification of $G_f$ (involving only values of $f$ at location $x$) and $\lambda_f$ its associated polar decomposition. This model
 can be written:
\begin{eqnarray}
&& \partial_t f^\varepsilon + \nabla_x \cdot  (A\, e_1 \, f^\varepsilon) = \frac{1}{\varepsilon} Q(f^\varepsilon) \label{eq:body_KM31_eps} \\
&& Q(f) =  \nabla_A \cdot \big( - (P_{T_A} \, \, F_f) \, f + \tau \, \nabla_A f \big)\label{eq:body_KM4_eps} 
\end{eqnarray}
with $F_f$ given by (\ref{eq:body_KM2_eps}), (\ref{eq:body_KM3_eps}). The von Mises distribution is now defined by
\begin{equation}
M_{\kappa \Lambda} (A) = \frac{1}{Z} \exp \big( \kappa \,  \Lambda \cdot A \big), 
\label{eq:vmf_body}
\end{equation}
where $\Lambda \cdot A$ is the matrix inner product of $\Lambda$ and $A$ defined above, $\kappa = \nu / \tau$ and $Z$ is a normalization constant only depending on $\kappa$. 
Then, $Q(f)$ can be written as 
\begin{eqnarray}
&& Q(f) = {\mathcal Q} (f;\lambda_f),
\label{eq:body_QQ} 
\end{eqnarray}
where $\lambda_f$ is given by (\ref{eq:body_KM2_eps}) and 
\begin{eqnarray}
&& {\mathcal Q} (f;\lambda)(A) = \tau \, \nabla_A \cdot \Big( M_{\kappa \lambda} (A) \nabla_A \big( \frac{f(A)}{M_{\kappa \lambda} (A)} \big) \Big). 
\label{eq:body_calQ} 
\end{eqnarray}
In the same way as before, as $\varepsilon \to 0$, $f^\varepsilon \to f^0$, where $f^0$ is a solution of $Q(f^0)=0$. This implies the existence of $\rho= \rho(x,t) \in [0,\infty)$ and $\lambda = \lambda(x,t) \in \mbox{SO}(3)$ such that 
\begin{equation}
f^0(x,A,t) = \rho(x,t) \, M_{\kappa \lambda(x,t)} (A).
\label{eq:vicsek_equi_body}
\end{equation}
Now, we define the GCI as follows:
\begin{definition}
For a body orientation given by the rotation matrix $\lambda \in \mbox{SO}(3)$, we define a GCI associated with $\lambda$ as a function $\psi(A) $ such that 
\begin{eqnarray}
\int {\mathcal Q} (f;\lambda)(A) \, \psi(A) \,  dA  =  0, \quad \forall f \, \mbox{ such that } \, P_{T_A} g_f =0. 
\label{eq:vel_calQ_body} 
\end{eqnarray}
\label{def:GCI_body}
\end{definition}

Up to now, the above body attitude alignment model could have been written in any dimension, i.e. for $A \in \mbox{SO}(d)$ for any dimension $d$. The following characterization of the set of GCI now requires the dimension $d$ to be equal to $3$. A characterization like this in the case of a general dimension $d$ is still an open problem. 
\begin{theorem}
The set ${\mathcal C}_\lambda$ of GCI associated to the body orientation given by the rotation matrix $\lambda \in \mbox{SO}(3)$ is a linear vector space of dimension $4$ expressed as follows: 
\begin{eqnarray}
{\mathcal C}_\lambda = \{ C + P \cdot (\lambda^T \, A) \, h(\lambda \cdot A) \, \, | \, \, C \in {\mathbb R}, \, \, P \in {\mathcal A} \}, 
\label{eq:GCI_space_boody} 
\end{eqnarray}
where ${\mathcal A}$ denotes the space of antisymmetric $3 \times 3$ matrices
and where $h$: $(0,\pi) \to {\mathbb R}$ is the unique solution of 
\begin{eqnarray}
&& \hspace{-1.5cm}
- \frac{d}{d \theta} \Big( \sin^2 (\theta/2) \, m(\theta) \, \frac{d}{d \theta} \big(\sin \theta \, h(\theta) \big) \Big) + \frac{1}{2} \, \sin \theta \, m(\theta)  \, h (\theta)\nonumber \\
&& \hspace{4.5cm}
 = - \sin^2 (\theta/2) \, \sin \theta \, m(\theta),
\label{eq:def_h_body} 
\end{eqnarray}
in the space 
\begin{eqnarray}
&& \hspace{-1cm}
H = \{ h: \, (0,\pi) \to {\mathbb R} \, \, | \nonumber \\
&& \hspace{1cm}
 \sin \theta \, h \in L^2(0,\pi), \, \,  \sin (\theta/2) \, \frac{d}{d \theta} (\sin \theta \, h) \in L^2(0,\pi) \}.
\label{eq:def_H_body} 
\end{eqnarray}
Here, we have denoted by  
$$ m(\theta) = \frac{1}{Z} \, \exp \big( \kappa \, (\frac{1}{2} + \cos \theta ) \big), $$
where $Z$ is the normalization constant involved in (\ref{eq:vmf_body})  
\label{thm:GCI_body}
\end{theorem}

Using this expression of the GCI and the same methodology as in the previous section, in \cite{Degond_etal_M3AS2016}, we have proved the following:
\begin{theorem} 
Suppose that the solution $f^\varepsilon$ of  (\ref{eq:body_KM1_eps}), (\ref{eq:body_KM2_eps}) has a limit $f^0$ when $\varepsilon \to 0$. Then, $f^0$ is given by (\ref{eq:vicsek_equi_body}) where $\kappa = \nu/\tau$ and the pair $(\rho,\lambda)$: $(x,t) \in {\mathbb R}^3 \times [0,\infty) \mapsto  (\rho,\lambda)(x,t) \in [0,\infty) \times \mbox{SO}(3)$ satisfies 
the following ``self-organized hydrodynamics for body attitude coordination'' (SOHB) model: 
\begin{eqnarray}
&& \hspace{-1cm}
\partial_t \rho + c_1 \nabla_x \cdot (\rho \, \lambda e_1) = 0, \label{eq:soh1_body} \\
&& \hspace{-1cm}
\rho \big( \partial_t \lambda + c_2  (\lambda e_1 \cdot \nabla_x) \lambda \big) \nonumber \\
&& \hspace{0cm} 
+  \Big[ (\lambda e_1) \times \big( c_3 \, \nabla_x \rho + c_4 \, \rho \, r_x(\lambda) \big) + c_4 \, \rho \, \delta_x(\lambda) \,  \lambda e_1 \Big]_\times \, \lambda = 0 , \label{eq:soh2_body} 
\end{eqnarray}
with the coefficients $c_1$ to $c_4$ depending on $\nu$ and $\tau$. The quantities $r_x(\lambda)$ and $\delta_x(\lambda)$ are given by: 
\begin{eqnarray}
\delta_x(\lambda) = \mbox{Tr} \{ {\mathcal D}_x(\lambda) \}, \quad r_x(\lambda) = {\mathcal D}_x(\lambda) - {\mathcal D}_x(\lambda)^T, 
\label{eq:delta_r} 
\end{eqnarray}
where ${\mathcal D}_x(\lambda)$ is the matrix defined, for any vector $w \in {\mathbb R}^3$, as follows: 
\begin{eqnarray}
(w \cdot \nabla_x) \lambda = [{\mathcal D}_x(\lambda) w]_\times \lambda. 
\label{eq:def_Dx} 
\end{eqnarray}
Here and above, for a vector $w \in {\mathbb R}^3$, we denote by $[w]_\times$ the antisymmetric matrix defined for any vector $z \in {\mathbb R}^3$ by
\begin{eqnarray}
[w]_\times z = w \times z, 
\label{eq:[]_times} 
\end{eqnarray}
where $\times$ denote the cross product of two vectors.
\label{thm_SOH_body}
\end{theorem}

We note that (\ref{eq:def_Dx} ) makes sense as $(w \cdot \nabla_x) \lambda$ belongs to the tangent space $T_\lambda$ of SO$(3)$ at $\lambda$ and $T_\lambda =\{ P \, \lambda \, | \, P \in {\mathcal A} \}$. So, there exists $u \in {\mathbb R}^3$ such that $(w \cdot \nabla_x) \lambda = [u]_\times \, \lambda$ and since $u$ depends linearly on $w$, there exists a matrix ${\mathcal D}_x(\lambda)$ such that $u = {\mathcal D}_x(\lambda) w$. The notation ${\mathcal D}_x(\lambda)$ recalls that the coefficients of this matrix are linear combinations of first order derivatives of $\lambda$. Using the exponential map, in the neighborhood of any point $x_0$, we can write (omitting the time-dependence) $ \lambda(x) = \exp \big( [b(x)]_\times \big) \lambda(x_0)$ where $b$ is a smooth function from a neighborhood of $x_0$ into ${\mathbb R}^3$. It is shown in \cite{Degond_etal_M3AS2016} that 
$$ \delta_x(\lambda) (x_0) = (\nabla_x \cdot b)(x_0), \quad r_x(\lambda) (x_0) = (\nabla_x \times b) (x_0), $$
and thus, $\delta_x(\lambda)$ and $r_x(\lambda)$ can be interpreted as local ``divergence'' and ``curl'' of the matrix field $\lambda$. We note that (\ref{eq:soh2_body}) equally makes sense. Indeed, the expression on the first line is a derivative of the rotation field $\lambda$ and should consequently belong to $T_{\lambda(x,t)}$. But the second line has precisely the required structure as it is the product of an antisymmetric matrix with $\lambda$. Eq. (\ref{eq:soh1_body}) is the continuity equation for the density of agents moving at bulk velocity $c_1 \, \lambda e_1$ so that $\lambda e_1$ describes the fluid direction of motion. Eq. (\ref{eq:soh2_body}) gives the evolution of $\lambda$. The first line describes transport at velocity $c_2 \, \lambda e_1$ and since $c_2 \not = c_1$, the transport of $\lambda$ occurs at a different speed from the transport of mass, as in the SOH model (\ref{eq:soh1}), (\ref{eq:soh2}). The second line describes how $\lambda$ evolves during its transport. The first term (proportional to $\nabla_x \rho$) is the action of the pressure gradient and has the effect of turning the direction of motion away from high density regions. The other two terms are specific to the body attitude alignment model and do not have their counterpart in the classical SOH model (\ref{eq:soh1}), (\ref{eq:soh2}). The expressions of the coefficients $c_2$ to $c_4$ involve moments of the function $h$ intervening in the expression of the GCI. The mathematical theory of the SOHB model is entirely open. We note that the above theory can be recast in the unitary quaternion framework, as done in \cite{Degond_etal_MMS17}.

\section{Phase transitions}
\label{sec:phasetrans}

\subsection{A Vicsek model exhibiting multiple equilibria}
\label{subsec:Vic_multiple}

Now, we go back to the Vicsek model of Section \ref{subsec:vicsek}. More precisely, we consider the kinetic model (\ref{eq:vicsek_KM1_eps})-(\ref{eq:vicsek_KM3_eps}) in the spatially homogeneous case (i.e. we drop all dependences and derivatives with respect to position $x$) and with $\varepsilon = 1$. However, we are interested in the case where the coefficients $\tau$ and $\nu$ are functions of $|j_f|$. More precisely, we consider the system
\begin{eqnarray}
&& \hspace{-1.4cm} 
\partial_t f (v,t) = Q(f) (v,t), \label{eq:vicsek_KM1_homo} \\
&& \hspace{-1.4cm} 
Q(f) (v,t) =  \nabla_v \cdot \big( - \nu(|j_f(t)|) \, (P_{v^\bot} \,  u_f(t)) \, f(v,t) + \tau(|j_f(t)|) \, \nabla_v f (v,t) \big), \label{eq:vicsek_KM2_homo} \\
&& \hspace{-1.4cm} 
u_f(t)=\frac{j_f(t)}{|j_f(t)|},  \quad j_f(t) = \int_{{\mathbb S}^{d-1}} f(w,t) \, w \, dw. \label{eq:vicsek_KM3_homo} 
\end{eqnarray}
For future usage, we introduce the function $ k(|j|) = \frac{\nu(|j|)}{\tau(|j|)}$, as well as $\Phi$ the primitive of $k$: $\Phi(r) = \int_0^r k(s) \, ds$. Introducing the free energy 
\begin{eqnarray}
&& \hspace{-1.4cm} 
{\mathcal F}(f) = \int_{{\mathbb S}^{d-1}} f(v) \, \log f(v) \, dv - \Phi(|j_f|), 
\label{eq:free_ener} 
\end{eqnarray}
we find the free energy dissipation inequality
\begin{eqnarray}
&& \hspace{-1.4cm} 
\frac{d}{dt} {\mathcal F}(f)(t) = - {\mathcal D}(f) (t), 
\label{eq:free_ener_dissip_1} \\
&& \hspace{-1.4cm} 
{\mathcal D}(f) (t) = \tau(|j_f(t)|) \int_{{\mathbb S}^{d-1}} f(v,t) \, \Big| \nabla_v \big( f(v,t) -  k(|j_f(t)|) \, (v \cdot u_f(t) ) \big) \Big|^2 \, . 
\label{eq:free_ener_dissip_2} 
\end{eqnarray}

In \cite{Degond_etal_arXiv:1304.2929} (see a special case in \cite{Degond_etal_JNonlinearSci13}), we first give the proof of the following 

\begin{theorem}
\label{theorem-existence-uniqueness}
Given an initial finite nonnegative measure~$f_0$ in the Sobolev space~$H^s({\mathbb S}^{d-1})$, there exists a unique weak solution~$f$ of~\eqref{eq:vicsek_KM1_homo}  such that~$f(0)=f_0$.
This solution is global in time.
Moreover,~$f\in C^1(\mathbb{R}^*_+,C^\infty({\mathbb S}^{d-1}))$, with~$f(v,t)>0$ for all positive~$t$. Furethermore, we have the following instantaneous regularity and uniform boundedness estimates (for~$m\in\mathbb{N}$, the constant~$C$ being independent of~$f_0$):
\[\|f(t)\|^2_{H^{s+m}}\leqslant C\left(1+\frac1{t^m}\right)\|f_0\|^2_{H^{s}}.\]
For these solutions, the density $\rho(t) = \int_{{\mathbb S}^{d-1}} f(v,t) \, dv$ is constant in time, i.e. $\rho(t) = \rho$, where $\rho = \int_{{\mathbb S}^{d-1}} f_0(v) \, dv$.
\end{theorem}

The equilibria, i.e. the solutions of $Q(f)=0$ are given by $\rho \, M_{\kappa u}$ where $\rho$ is the initial density as defined in Theorem \ref{theorem-existence-uniqueness} and $M_{\kappa u}$ is still the von Mises Fisher distribution (\ref{eq:vmf}) with arbitrary value of $u \in {\mathbb S}^{d-1}$. However, now the value of $\kappa$ is found by the resolution of a fixed-point equation (the consistency condition)
\begin{eqnarray}
&& \hspace{-1.4cm} 
\kappa = k(|j_{\rho M_{\kappa u}}|). 
\label{eq:fixed_point_kappa} 
\end{eqnarray}
This equation can be recast by noting that $|j_{\rho M_{\kappa u}}| = \rho \, c_1(\kappa)$ where $c_1(\kappa)$ is the order parameter (\ref{eq:order_param}). Assuming that the function $k$: $|j| \in [0,\infty) \mapsto k(|j|) \in [0,\infty)$ is strictly increasing and surjective, we can define its inverse $\iota$: $\kappa \in [0, \infty) \mapsto \iota(\kappa) \in [0,\infty)$. This assumption may be seen as restrictive, but it is easy to remove it at the expense of more technicalities, which we want to avoid in this presentation. As by definition $\iota(k(|j|)) = |j|$, applying the function $\iota$ to (\ref{eq:fixed_point_kappa}), we can recast it in 
\begin{eqnarray}
&& \hspace{-1.4cm} 
\mbox{either } \quad  \kappa = 0 \quad \mbox{ or } \quad  \frac{\iota(\kappa)}{c_1(\kappa)} = \rho . 
\label{eq:fixed_point_kappa_2} 
\end{eqnarray}
Note that for $\kappa = 0$, the von Mises distribution is the uniform distribution on the sphere. We will call the corresponding equilibrium, ``isotropic equilibrium''. Any von Mises distribution with $\kappa > 0$ will be called a ``non-isotropic equilibrium''. For a given $\kappa >0$, the von Mises equilibria $\rho \, M_{\kappa u}$ form a manifold diffeomorphically parametrized by $u \in {\mathbb S}^{d-1}$. Both $\iota$ and $c_1$ are increasing functions of $\kappa$ so the ratio $\frac{\iota(\kappa)}{c_1(\kappa)}$ has no defined monotonicity a priori. For a given $\rho$ the number of solutions $\kappa$ of (\ref{eq:fixed_point_kappa_2}) depends on the particular choice of the function $k$. However, we can state the following proposition:

\begin{proposition}
\label{prop-two-thresholds}
Let~$\rho>0$.
We define
\begin{align}
\label{def-rho-c}
\rho_c=\lim_{\kappa\to 0} \frac{\iota(\kappa)}{c_1(\kappa)}, \quad \rho_*=\inf_{\kappa\in(0,\infty)} \frac{\iota(\kappa)}{c_1(\kappa)},
\end{align}
where~$\rho_c>0$ may be equal to~$+\infty$.
Then we have~$\rho_c\geqslant\rho_*$, and
\begin{itemize}
\item[(i)] If~$\rho<\rho_*$, the only solution to (\ref{eq:fixed_point_kappa_2}) is~$\kappa=0$ and the only equilibrium with total mass~$\rho$ is the uniform distribution~$f=\rho$.
\item[(ii)] If~$\rho>\rho_*$, there exists at least one positive solution~$\kappa>0$ to (\ref{eq:fixed_point_kappa_2}). It corresponds to a family~$\{\rho M_{\kappa u}, \, u \in {\mathbb S^{d-1}}\}$ of non-isotropic von Mises equilibria.
\item[(iii)] The number of families of nonisotropic equilibria changes as~$\rho$ crosses the threshold~$\rho_c$. Under regularity and non-degeneracy hypotheses, in a neighborhood of~$\rho_c$, this number is even when~$\rho<\rho_c$ and odd when~$\rho>\rho_c$.
\end{itemize}
\end{proposition}

Now, the key question is the stability of these equilibria. A first general result can be established thanks to the La Salle principle: 

\begin{proposition}
\label{prop-lasalle-refined}
Let~$f_0$ be a positive measure on the sphere~${\mathbb S}^{d-1}$, with mass~$\rho$, and $f(t)$ the associated solution to (\ref{eq:vicsek_KM1_homo}). If no open interval is included in the set~$\{\kappa \in [0,\infty) \, | \, \rho c(\kappa)=\iota(\kappa)\}$, then there exists a solution~$\kappa_\infty$ to (\ref{eq:fixed_point_kappa_2}) such that:
\begin{gather}
\label{eq-limJ}
\lim_{t\to\infty} |j_f(t)|=\rho c(\kappa_\infty)
\intertext{and}
\label{eq-limf}
\forall s\in\mathbb{R}, \lim_{t\to\infty}\,\|f(t)-\rho M_{\kappa_\infty u_f(t)}\|_{H^s}=0.
\end{gather}
\end{proposition}
In other words, under these conditions, the family of equilibria $\{ \rho M_{\kappa_\infty u} \, | \, u \in {\mathbb S}^{d-1} \}$ is an $\omega$-limit set of the trajectories of (\ref{eq:vicsek_KM1_homo}). Now, we study separately the stability of the isotropic and non-isotropic equilibria.

\subsection{Stability of the isotropic equilibria}
\label{subsec:stab_iso}

For the isotropic equilibria, we have the following two propositions: 

\begin{proposition}
\label{prop-unstability-uniform}
Let~$f(t)$ be the solution to (\ref{eq:vicsek_KM1_homo}) associated with initial condition $f_0$ of  mass~$\rho$. If~$\rho>\rho_c$, and if~$j_{f_0}\neq0$, then we cannot have~$\kappa_\infty=0$ in Proposition~\ref{prop-lasalle-refined}.
\end{proposition}

\begin{proposition}
\label{prop-stability-uniform}
Suppose that~$\rho<\rho_c$.
We define
\begin{equation}
\lambda=(n-1)\tau_0(1-\frac{\rho}{\rho_c})>0.
\label{eq:decayrate_iso}
\end{equation}
Let~$f_0$ be an initial condition with mass~$\rho$, and~$f$ the corresponding solution to (\ref{eq:vicsek_KM1_homo}). There exists~$\delta>0$ independent of~$f_0$ such that if~$\|f_0-\rho\|_{H^s}<\delta$, then for all~$t\geqslant0$
\[\|f(t)-\rho\|_{H^s}\leqslant\frac{\|f_0-\rho\|_{H^s}}{1-\frac{1}{\delta}\|f_0-\rho\|_{H^s}}e^{-\lambda t}.\]
\end{proposition}

Prop. \ref{prop-unstability-uniform} implies the instability of the uniform equilibria for $\rho>\rho_c$ (provided the initial current $j_{f_0}$ does not vanish) as the $\omega$-limit set of the trajectories consists of non-isotropic equilibria. Prop. \ref{prop-stability-uniform} shows the stability of the uniform equilibria for $\rho<\rho_c$ in any $H^s$ norm with exponential decay rate given by (\ref{eq:decayrate_iso}). We stress that these are fully nonlinear stability/instability results.

\subsection{Stability of the non-isotropic equilibria}
\label{subsec:stab_aniso}

Let~$\kappa>0$ and $\rho>0$ be such that~$\kappa$ is a solution to (\ref{eq:fixed_point_kappa_2}). In addition to the hypotheses made so far on $k$, we assume that $k$ is differentiable, with its derivative~$k'$ being itself Lipschitz. The following result shows that the stability or instability of the non-isotropic equilibria is determined by whether the function $\kappa \mapsto \frac{\iota(\kappa)}{c_1(\kappa)}$ is strictly increasing or decreasing.  

\begin{proposition}
\label{prop-unstability-stability-anisotropic}
Let~$\kappa>0$ and~$\rho=\frac{\iota(\kappa)}{c_1(\kappa)}$.
We denote by~$\mathcal F_\kappa$ the value of~$\mathcal F(\rho M_{\kappa u})$ (independent of~$u\in {\mathbb S}^{d-1}$).

\begin{itemize}
\item[(i)] Suppose~$(\frac{\iota}{c_1})'(\kappa)<0$.
Then any equilibrium of the form~$\rho M_{\kappa u}$ is unstable, in the following sense:
in any neighborhood of~$\rho M_{\kappa u}$, there exists an initial condition~$f_0$ such that~$\mathcal F(f_0)<\mathcal F_\kappa$.
Consequently, in that case, we cannot have~$\kappa_\infty=\kappa$ in Proposition~\ref{prop-lasalle-refined}.
\item[(ii)] Suppose~$(\frac{\iota}{c_1})'(\kappa)>0$.
Then the family of equilibria~$\{\rho M_{\kappa u}, u \in {\mathbb S}^{d-1}\}$ is stable, in the following sense: for all~$K>0$ and~$s>\frac{n-1}2$, there exists~$\delta>0$ and~$C$ such that for all~$f_0$ with mass~$\rho$ and with~$\|f_0\|_{H^s}\leqslant K$, if~$\|f_0-\rho M_{\kappa u}\|_{L^2}\leqslant\delta$ for some~$u \in {\mathbb S}^{d-1}$, then for all~$t\geqslant0$, we have
\begin{gather*}
\mathcal F(f)\geqslant\mathcal F_\kappa,\\
\|f-\rho M_{\kappa u_f}\|_{L^2}\leqslant C\|f_0-\rho M_{\kappa u_{f_0}}\|_{L^2}.
\end{gather*}
\end{itemize}
\end{proposition}

Note that the marginal case $(\frac{\iota}{c_1})'(\kappa)=0$ is not covered by the above theorem and is still an open problem. In the stable case, the following proposition provides the rate of decay to an element of the same family of equilibria:

\begin{theorem}
\label{thm-strong-stability-anisotropic}
Suppose~$(\frac{\iota}{c_1})'(\kappa)>0$.
Then, for all~$s>\frac{n-1}2$, there exist cons\-tants~$\delta>0$ and~$C>0$ such that for any~$f_0$ with mass~$\rho$ satisfying~$\|f_0-\rho M_{\kappa u}\|_{H^s}<\delta$ for some~$u \in {\mathbb S}^{d-1}$, there exists~$u_\infty \in {\mathbb S}^{d-1}$ such that
\[\|f-\rho M_{\kappa u_\infty}\|_{H^s}\leqslant C\|f_0-\rho M_{\kappa u}\|_{H^s} \, e^{-\lambda t},\]
where the rate $\lambda$ is given by
\begin{equation}
\label{def-lambda}
\lambda=\frac{c_1(\kappa) \, \tau(\iota(\kappa))}{\iota'(\kappa)} \, \Lambda_\kappa \, \big(\frac{\iota}{c_1} \big)'(\kappa).
\end{equation}
The constant~$\Lambda_\kappa$ is the best constant for the following weighted Poincaré inequality (see the appendix of~\cite{Degond_etal_JNonlinearSci13}):
\begin{equation}
\label{poincare-lambda}
\langle|\nabla_\omega g|^2\rangle_{M}\geqslant\Lambda_\kappa\langle(g-\langle g\rangle_{M})^2\rangle_{M}, 
\end{equation}
where we have writen~$\langle g \rangle_M$ for~$\int_\mathbb{S}g(v)M_{\kappa u} (v) \, dv$.
\end{theorem}

\section{Conclusion}
\label{sec:conclusion}

In this short overview, we have surveyed some of the mathematical questions posed by collective dynamics and self-organization. We have particularly focused on two specific problems: the derivation of macroscopic models and the study of phase transitions. There are of course many other fascinating challenges posed by self-organized systems. These have shown to be an inexhaustible source of problems for mathematicians and a drive for the invention of new mathematical concepts.




\vspace{-0.6cm}

\begin{thebibliography}{99}

\vspace{-0.2cm}

\setlength{\itemsep}{-3pt}

\bibitem{Aoki_BullJapSocSciFish92} I. Aoki, A simulation study on the schooling mechanism in fish, Bulletin of the Japan Society of Scientific Fisheries, 48 (1982) 1081-1088.

\bibitem{Bak_etal_PRL87}
P. Bak, C. Tang, K. Wiesenfeld, Self-organized criticality: an explanation of $1/f$ noise, Phys. Rev. Lett., 59 (1987) 381-384. 

\bibitem{Barbaro_Degond_DCDSB13} A. Barbaro, P. Degond, Phase transition and diffusion among socially interacting self-propelled agents, Discrete Contin. Dyn. Syst. Ser. B, to appear. 

\bibitem{Bazazi_etal_CurrBiol08} S. Bazazi et al, Collective Motion and Cannibalism in Locust Migratory Bands, Current Biology 18 (2008) 735-739. 

\bibitem{Berthelin_etal_ARMA08} F. Berthelin, P. Degond, M. Delitala, M. Rascle,  A model for the formation and evolution of traffic jams, Arch. Rat. Mech. Anal., 187 (2008) 185-220. 

\bibitem{Bertin_etal_JPhysA09} E.~Bertin, M.~Droz and G.~Gr\'egoire, Hydrodynamic equations for self-propelled particles, J. Phys. A: Math. Theor.  42 (2009) 445001.

\bibitem{Boissard_etal_JMB13}
E. Boissard, P. Degond, S. Motsch, Trail formation based on directed pheromone deposition, J. Math. Biol., 66 (2013) 1267-1301.

\bibitem{Bolley_etal_AML11}  F. Bolley, J. A. Ca\~nizo, J. A. Carrillo, Mean-field limit for the stochastic Vicsek model, Appl. Math. Lett., 25 (2011) 339-343.

\bibitem{Carlen_etal_M3AS13}
E. Carlen, P. Degond, and B Wennberg, Kinetic limits for pair-interaction driven master equations and biological swarm models, Math. Models Methods Appl. Sci., 23 (2013) 1339-1376. 

\bibitem{Carrillo_etal_SIMA10} J. A. Carrillo et al, Asymptotic Flocking Dynamics for the kinetic Cucker-Smale model, SIAM J. Math. Anal., 42 (2010) 218-236.

\bibitem{Chuang_etal_PhysicaD07}  Y-L. Chuang et al, State transitions and the continuum limit for a 2D interacting, self-propelled particle system, Physica D, 232 (2007) 33-47. 

\bibitem{Couzin_etal_JTB02}  
I. D. Couzin et al, Collective Memory and Spatial Sorting in Animal Groups, J. theor. Biol., 218 (2002) 1-11. 

\bibitem{Creppy_etal_Interface16}
A. Creppy et al, Symmetry-breaking phase-transitions in highly concentrated semen, Journal of the Royal Society Interface, 13 (2016), p. 20160575. 

\bibitem{Cucker_Smale_IEEETransAutCont07} F. Cucker, S. Smale, Emergent behavior in flocks, IEEE Transactions on Automatic Control, 52 (2007) 852-862.

\bibitem{Czirok_etal_PRE96} A. Czir\`ok, E. Ben-Jacob, I. Cohen, T. Vicsek, Formation of complex bacterial colonies via self-generated vortices, Phys. Rev. E, 54 (1996) 1791-18091.

\bibitem{Degond_etal_JNonlinearSci13}
P. Degond, A. Frouvelle, J-G. Liu, Macroscopic limits and phase transition in a system of self-propelled particles, J. Nonlinear Sci., 23 (2013) 427-456.

\bibitem{Degond_etal_arXiv:1304.2929}
P. Degond, A. Frouvelle, J-G. Liu, Phase transitions, hysteresis, and hyperbolicity for self-organized alignment dynamics, Arch. Ration. Mech. Anal.,  216 (2015), pp 63-115.

\bibitem{Degond_etal_M3AS2016}
P. Degond, A. Frouvelle, S. Merino-Aceituno, A new flocking model through body attitude coordination, Math. Models Methods Appl. Sci.  27 (2017) 1005-1049.

\bibitem{Degond_etal_MMS17}
P. Degond, A. Frouvelle, S. Merino-Aceituno, A. Trescases, Quaternions in collective dynamics, Multiscale Model. Simul., to appear. arXiv:1701.01166

\bibitem{Degond_Hua_JCP13} P. Degond, J. Hua, Self-Organized Hydrodynamics with congestion and path formation in crowds, J. Comput. Phys., 237 (2013) 299-319.

\bibitem{Degond_etal_JCP11} P. Degond, J. Hua, L. Navoret, Numerical simulations of the Euler system with congestion constraint, J. Comput. Phys., 230 (2011) 8057-8088.

\bibitem{Degond_etal_MAA13}
P. Degond et al, Hydrodynamic models of self-organized dynamics: derivation and existence theory, Methods Appl. Anal., 20 (2013) 089-114.

\bibitem{Degond_etal_DCDSB16}
P. Degond, A. Manhart, H. Yu, A continuum model for nematic alignment of self-propelled particles,  DCDS B, 22 (2017) 1295-1327. 

\bibitem{Degond_Motsch_M3AS08}
P. Degond, S. Motsch,  Continuum limit of self-driven particles with orientation interaction, Math. Models Methods Appl. Sci., 18 Suppl. (2008) 1193-1215. 

\bibitem{Degond_Motsch_JSP08} P. Degond, S. Motsch,  Large scale dynamics of the Persistent Turning Walker model of fish behavior, J. Stat. Phys.,  131 (2008) 989-1021.

\bibitem{Degond_Motsch_JSP11} P. Degond, S. Motsch, A macroscopic model for a system of swarming agents using curvature control, J. Stat. Phys., 143 (2011) 685-714

\bibitem{Degond_etal_JSP10}
P. Degond et al, Congestion in a macroscopic model of self-driven particles modeling gregariousness, J. Stat. Phys., 138 (2010) 85-125.

\bibitem{Domeier_Colin_BullMarSci97} M. L. Domeier, P. L. Colin, Tropical reef fish spawning aggregations: defined and reviewed, Bulletin of Marine Science, 60 (1997) 698-726. 

\bibitem{Figalli_etal_arxiv15}
A. Figalli, M-J. Kang, J. Morales, Global well-posedness of the spatially homogeneous Kolmogorov-Vicsek model as a gradient flow, arXiv:1509.02599.

\bibitem{Frouvelle_M3AS12} A. Frouvelle, A continuum model for alignment of self-propelled particles with anisotropy and density-dependent parameters, Math. Mod. Meth. Appl. Sci., 22 (2012) 1250011 (40 p.).

\bibitem{Frouvelle_Liu_SIMA12} A. Frouvelle, J.-G. Liu, Dynamics in a kinetic model of oriented particles with phase transition, SIAM J. Math. Anal., 44 (2012) 791-826. 

\bibitem{Gallagher_etal_EMS13}
I. Gallagher, L. Saint-Raymond, B. Texier, From Newton to Boltzmann: hard spheres and short-range potentials. European math. soc., 2013.

\bibitem{Gamba_Kang_ARMA16}
I. M. Gamba,  M-J. Kang,  Global weak solution of the Kolmogorov-Fokker-Planck type equation with orientational interaction, Arch. Rat. Mech. Anal. 222 (2016) 317-342.


\bibitem{Gautrais_etal_JMB09} J. Gautrais et al, Analyzing fish movement as a persistent turning walker, J. Math. Biol., 58 (2009) 429-445.

\bibitem{Gautrais_etal_PlosCB12} J. Gautrais et al, Deciphering interactions in moving animal groups. Plos Comput. Biol., 8 (2012) e1002678.

\bibitem{Ginelli_etal_PRL10}
F. Ginelli, F. Peruani, M. B\"ar, H. Chat\'e, Large-scale collective properties of self-propelled rods, Phys. Rev. Lett. 104 (2010) 184502.

\bibitem{Ha_Liu_CMS09} S. -Y. Ha, J.-G. Liu, A simple proof of the Cucker-Smale flocking dynamics and mean-field limit, Commun. Math. Sci., 7 (2009) 297-325.

\bibitem{Ha_Tadmor_KRM08} S.-Y. Ha, E. Tadmor, From particle to kinetic and hydrodynamic descriptions of flocking, Kinetic and Related Models, 1 (2008) 415-435.

\bibitem{Haskovec_etal_NonlinAnalTMA16}
J. Haskovec et al, Notes on a PDE system for biological network formation, Nonlinear Anal. 138 (2016) 127-155.

\bibitem{Helbing_Farkas_Vicsek_PRL00}
D. Helbing, I. J. Farkas, T. Vicsek, Freezing by heating in a driven mesoscopic system, Phys. Rev. Lett. 84 (2000) 1240.

\bibitem{Hsu_AMS02} E.~P. Hsu, Stochastic Analysis on Manifolds, Graduate Series in Mathematics, American Mathematical Society, 2002.

\bibitem{Jiang_etal_arxiv15}
N. Jiang, L. Xiong, T-F. Zhang, Hydrodynamic limits of the kinetic self-organized models, 
SIAM J. Math. Anal. 48 (2016) 3383-3411.

\bibitem{Khuong_etal_ECAL11} A. Khuong et al, A computational model of ant nest morphogenesis, in "Advances in Artificial Life, ECAL 2011, MIT Press, 2011, pp. 404-411. 

\bibitem{Lanford_76} O. E. Lanford, III. On a derivation of the Boltzmann equation. In Ast\'erisque No. 40,  Soc. Math. France, Paris, 1976, pp. 117-137. 

\bibitem{Leroy_etal_BMB17}
M. Leroy-Ler\^etre et al, Are tumor cell lineages solely shaped by mechanical forces ? Bull. Math. Biol., 79 (2017) 2356-2393.

\bibitem{Lukeman_etal_PNAS10}
R. Lukemana, Y.-X. Li, L. Edelstein-Keshet, Inferring individual rules from collective behavior, Proc. Natl. Acad. Sci. USA 107 (2010), 12576-12580.

\bibitem{Mischler_Mouhot_InvMath13}
S. Mischler, C. Mouhot, Kac's Program in Kinetic Theory, Invent. Math., 193 (2013) 1-147,

\bibitem{Motsch_Tadmor_JSP11} S. Motsch, E. Tadmor, A new model for self-organized dynamics and its flocking behavior, J. Stat. Phys., 144 (2011) 923-947.

\bibitem{Moussaid_etal_PlosCB12}
M. Moussaïd et al, Traffic Instabilities in Self-organized Pedestrian Crowds, PLoS Computational Biology, 8 (2012) e1002442

\bibitem{Perthame_etal_ARMA14}
B. Perthame, F. Quir\`os, J. L.  V\`azquez, The Hele-Shaw asymptotics for mechanical models of tumor growth, Arch. Ration. Mech. Anal. 212 (2014) 93-127.

\bibitem{Peurichard_etal_JTB17}
D. Peurichard et al, Simple mechanical cues could explain adipose tissue morphology, J. Theoret. Biol., 429 (2017), 61-81.

\bibitem{Poujade_etal_PNAS07}
M. Poujade et al, Collective migration of an epithelial monolayer in response to a model wound, Proc. Natl. Acad. Sci. USA 104 (2007), 15988-15993.

\bibitem{Shraiman_PNAS05}
B. I. Shraiman, Mechanical feedback as a possible regulator of tissue growth, Proc. Natl. Acad. Sci. USA 102 (2005), 3318-3323.

\bibitem{Shen_SIAP07} J. Shen, Cucker-Smale flocking under hierarchical leadership, SIAM J. Appl. Math., 58 (2007) 694-719. 

\bibitem{Toner_etal_PRE98} 
J. Toner, Y. Tu, Flocks, herds, and schools: A quantitative theory of flocking, Phys. Rev. E 58 (1998), 4828.

\bibitem{Vicsek_etal_PRL95}
T. Vicsek et al, Novel type of phase transition in a system of self-driven particles,
Phys. Rev. Lett., 75 (1995) 1226-1229.

\bibitem{Vicsek_Zafeiris_PhysRep12} 
T. Vicsek, A. Zafeiris, Collective motion, Phys. Rep., 517 (2012) 71-140.

\end{thebibliography}
\end{document}